\documentclass[a4paper,prd,showpacs,aps,amsmath,amssymb,preprint]{revtex4}
\usepackage{graphicx}
\usepackage{graphics}
\usepackage{amsmath}
\usepackage{dcolumn}
\usepackage{amssymb}
\usepackage{bm}

\begin{document}
\title{On Bargmann Representations of Wigner Function}
\author{Fernando Parisio}
\affiliation{Departamento de F\'{\i}sica, Universidade Federal de Pernambuco, 50670-901, Recife, PE, Brazil}

\begin{abstract}
By using the localized character of canonical coherent states, we give a straightforward derivation of the Bargmann integral representation of Wigner function ($W$). A non-integral representation is presented in terms of a quadratic form $W \propto {\bf V}^\dagger {\bf F}{\bf V}$, where ${\bf F}$ is a self-adjoint matrix whose entries are tabulated functions and ${\bf V}$ is a vector depending in a simple recursive way on the derivatives of the Bargmann function. Such a representation may be of use in numerical computations. We discuss a relation involving the geometry of Wigner function and the spacial uncertainty of the coherent state basis we use to represent it.
\end{abstract}

\maketitle

\section{Introduction}
The Heisenberg uncertainty principle assures that no function of the canonically conjugated pair $(q,p)$ can be
defined in a way to be interpreted as a genuine probability density in phase-space. Despite this fact, many phase-space representations of quantum mechanics have been developed and demonstrated to be powerfull tools in different fields, such as semiclassical limit \cite{rivas,dittrich,brac,tos} and quantum optics \cite{nielsen,nee} .

In 1932 Wigner \cite{wig} introduced his famous {\it quasi-probability} function $W$, so named for its possible negativity in some regions of phase-space. For a pure ensemble with density operator $ \hat{\rho}=|\psi \rangle \langle \psi |$, $W$ is usually presented as an integral in configuration space:
\begin{equation}
\label{wig0}
W(q,p)=\frac{1}{2 \pi \hslash}\int {\rm d}y \, \langle q+y/2|\psi \rangle \langle \psi | q-y/2 \rangle \, e^{-i py/\hslash} \;.
\end{equation}
Among other appealing properties, the above function yields the correct marginal probabilities: integration of 
$W$ in the variable $p$ gives the position probability density, and conversely, integration in the $q$-axis leads to the probability density in momentum space. The Wigner function contains all the information on the ensemble, providing an alternative formalism for quantum mechanics. In the case of mixed ensembles it enables the calculation of quantum averages in a classical way.

Considering that the simplest way to address quantum mechanics in phase-space is, arguably, 
by employing coherent states, it is natural to ask how the Wigner function is related to this representation. Not surprisingly, this connection has been partially studied by Cahill and Glauber \cite{cg}, and W\"unsche \cite{wun1,wun2}, on very formal grounds. We shall be concerned with the slightly different, but equivalent question of how the Wigner function is related to the analytical representation associated to the coherent states (the so-called Bargmann representation \cite{bar}).

Our purpose in
this work is, firstly, to use the fact that coherent states are maximally localized structures in phase-space to give a simple derivation of the Bargmann integral representation of Wigner function (section III). We believe the method we use to obtain this result is simpler than the previous ones \cite{cg,wun1,wun2}. It provides more physical insight and leads to the non-integral representation given in section IV in a natural way. These discrete representations have been often used in numerical evaluations of Wigner function in theoretical \cite{moya} and experimental situations, e. g., in the reconstruction of motional states of trapped atoms and ions \cite{leib, bode}. In section V we discuss a relation involving the geometry of Wigner function and the spacial uncertainty of the coherent state basis we use to represent it. Our final comments are outlined in section VI.

\section{Preliminary Definitions}
Canonical coherent states \cite{klauder,Perelomov} are defined in terms of number states associated to quantum harmonic oscillators:
\begin{equation}
| z' \rangle = e^{-{z'}^*z'/2}\sum_n \frac{{z'}^n}{\sqrt{n!}} |n \rangle\; ,
\end{equation}
where $\sqrt{2}\,z'=(q'/b+ibp'/\hslash)$ is a complex label, $q'$ and $p'$ are the mean values of the related quantum operators in the state $| z' \rangle$, and $b$ is a positive arbitrary constant. Such states constitute an over-complete basis \cite{cahill} of the Hilbert space, and therefore, can be used to express the resolution of unit:
\begin{equation}
\nonumber
\hat{I}=\int \frac{{\rm d}^2 z'}{\pi} | z' \rangle \langle z'| \; ,  
\end{equation}
where ${\rm d}^2 z'/\pi = {\rm d}q' {\rm d}p'/2 \pi \hslash$. We recall that in the position representation we have
\begin{equation}
\label{gauss}
\langle y| z' \rangle = \pi^{-1/4}b^{-1/2}\exp\left\{-\frac{1}{2}(y/b-\sqrt{2}z')^2+\frac{z'}{2}(z'-{z'}^*)\right\}\; ,
\end{equation}
from which it is clear that the parameter $b$ is related to the uncertainty in position of the state $| z' \rangle$ ($b=\sqrt{2}\Delta q'$). This constant can be freely chosen. Note, however, that the minimum uncertainty relation $\Delta q' \Delta p' =\hslash/2$ must be satisfied (implying $b=\sqrt{2} \hslash/\Delta p'$).

\section{Integral Representation Revisited}
In order to define the Wigner function as a phase-space integral we start by conveniently inserting coherent-state unit operators in expression (\ref{wig0})
\begin{equation}
W(q,p)=\frac{1}{2 \pi \hslash}\int \frac{{\rm d}^2 z'}{\pi} \, \frac{{\rm d}^2 z''}{\pi} \, \langle z''| \psi \rangle \langle \psi| z' \rangle 
\int {\rm d}y \, \langle q+y/2| z'' \rangle \langle z' | q-y/2 \rangle \, e^{-i py/\hslash} \;.
\end{equation}
Since $\langle q+y/2| z'' \rangle \langle z' | q-y/2 \rangle $ is a Gaussian function of $y$ (see equation (\ref{gauss})) the integration in this variable readily gives
\begin{equation}
W(q,p)=W(z,z^*)=\frac{e^{-2z^*z}}{\pi \hslash}\int \frac{{\rm d}^2 z'}{\pi} \, \frac{{\rm d}^2 z''}{\pi} \, \langle z''| \psi \rangle \langle \psi| z' \rangle 
e^{2{z'}^*z+2z^*z''-{z''}^*z''/2-{z'}^*z'/2-{z'}^*z''} \; .
\end{equation}
In order to better explore the properties of coherent-states we change variables as follows: $z''=w$ and $z'=w+\delta w$. Since the Jacobian determinant is unitary one gets
\begin{eqnarray}
\nonumber
W(z,z^*)=\frac{e^{-2z^*z}}{\pi \hslash}\int \frac{{\rm d}^2 w}{\pi} \, \frac{{\rm d}^2 (\delta w)}{\pi} \, \langle w| \psi \rangle \langle \psi| w+\delta w \rangle  \\ 
\exp\{-2w^*w+2z^*w+2zw^*-\delta w^*\delta w/2-(3w/2-2z)\delta w^*-w^*\delta w/2\} \; .
\label{wigcor}
\end{eqnarray}
Note that, due to the minimum uncertainty character of coherent states, the integration in $w$ and $w^*$ involving $\langle w| \psi \rangle\langle \psi| w+\delta w \rangle$, with $\psi$ being a $L^2$ function, falls-off in a Gaussian-like way for $\delta w \ne 0$. Therefore, it is safe to make an expansion around this point. However, before proceding to such an expansion, we recall that $\langle \psi| w+\delta w\rangle$ is not an analytical function of $w+\delta w$, since there is a (trivial) dependence on $w^*+\delta w^*$, so that $\langle \psi| w+\delta w\rangle = f_{\psi}(w+\delta w,w^*+\delta w^*)$. Thus, the referred expansion, in the coherent state representation, amounts for a two variable Taylor series. This unnecessary complication can be avoided in the Bargmann representation, defined as  $ f(w) \equiv \exp(w^*w/2)\langle \psi| w\rangle$, which provides a description based on entire functions \cite{bar}. Re-writing the Wigner function in this formalism we get
\begin{equation}
W(z,z^*)=\frac{e^{-2z^*z}}{\pi \hslash}\int \frac{{\rm d}^2 w}{\pi} \, \frac{{\rm d}^2 (\delta w)}{\pi} \, f^*(w)f(w+\delta w)\, e^{-3w^*w+2z^*w+2zw^*}
e^{-\delta w^*\delta w-2(w-z)\delta w^*-w^*\delta w} \; .
\end{equation}
Since $f(w+\delta w)$ is an analytic function of its argument one can proceed a single variable Taylor expansion:
$f(w+\delta w)=\sum_{n=0}^{\infty} \, \frac{1}{n!}\frac{{\rm d}^n }{{\rm d} w^n}f(w) \,\delta w^n$.
We obtain
\begin{eqnarray}
\nonumber
W(z,z^*)=\frac{e^{-2z^*z}}{\pi \hslash}\int \frac{{\rm d}^2 w}{\pi} \, f^*(w)\, e^{-3w^*w+2z^*w+2zw^*} \sum_{n=0}^{\infty} \, \frac{1}{n!}\frac{{\rm d}^n }{{\rm d} w^n}f(w) \\ 
\times \int \frac{{\rm d}^2 (\delta w)}{\pi} \,\delta w^n e^{-\delta w^*\delta w-2(w-z)\delta w^*-w^*\delta w} \; .
\end{eqnarray}
The integral in $\delta w$ and $\delta w^*$ is given by 
\begin{equation}
(-1)^n \frac{\partial^n}{\partial {w^*}^n} \int \frac{{\rm d}^2 (\delta w)}{\pi} \, e^{-\delta w^*\delta w-2(w-z)\delta w^*-w^*\delta w} = 2^n(z-w)^n\, e^{-2w^*(z-w)}\; ,
\end{equation}
thus
\begin{equation}
W(z,z^*)=\frac{e^{-2z^*z}}{\pi \hslash}\int \frac{{\rm d}^2 w}{\pi} \, f^*(w)\, e^{-w^*w+2z^*w} \sum_{n=0}^{\infty} \, \frac{2^n}{n!}(z-w)^n\frac{{\rm d}^n }{{\rm d} w^n}f(w) \; .
\end{equation}
By writing $f(w)=\sum_{k=0}^\infty a_k w^k$ we have $\frac{{\rm d}^n }{{\rm d} w^n}f(w)=\sum_{k=0}^\infty a_k\frac{k!}{(k-n)!} w^{k-n}$, where we used the fact that $1/|\Gamma(-N)| \rightarrow 0$ for $N=0,1,2,...$ to extend the second sum in $k$ from $(n, \infty)$ to $(0, \infty)$. This leads to
\begin{eqnarray}
\nonumber
\sum_{n=0}^{\infty} \, \frac{2^n}{n!}(z-w)^n\frac{{\rm d}^n }{{\rm d} w^n}f(w)=\sum_{k=0}^\infty a_k \sum_{n=0}^{\infty} \, \frac{k!}{n!(k-n)!} \,(2z -2w)^n w^{k-n} \\ \nonumber
=\sum_{k=0}^\infty a_k (2z -w)^k=f(2z-w)\; ,
\end{eqnarray} 
where we used the binomial expansion. We finally get
\begin{equation}
\label{wigph}
W(z,z^*)=\frac{e^{-2z^*z}}{\pi \hslash}\int \frac{{\rm d}^2 w}{\pi} \, f^*(w)f(2z-w)\, e^{-w^*w+2z^*w} \; ,
\end{equation}
which can be written in a more symmetrical way as:
\begin{equation}
\label{wigsim}
W(z,z^*)=\frac{e^{-z^*z}}{4 \pi \hslash}\int \frac{{\rm d}^2 w}{\pi} \, f^*(z+w/2)f(z-w/2)\, e^{-w^*w/4+z^*w/2-zw^*/2} \; ,
\end{equation}
where $f^*(z+w/2)=\exp\{ (z+w/2)(z^*+w^*/2)/2\} \langle z+w/2| \psi \rangle$ and $f(z-w/2)=\exp\{ (z-w/2)(z^*-w^*/2)/2\} \langle \psi | z-w/2\rangle$. This expression is the phase-space analogous of equation (\ref{wig0}). It is clear that in the case of a mixed ensemble of states $| \psi_i \rangle$ with statistical weights $p_i$ ($ \hat{\rho}=\sum_i p_i |\psi_i \rangle \langle \psi_i |$) we have $W=\sum_i p_i W_{i}$, where each $ W_{i}$ is given by equation (\ref{wigsim}) 

\section{Non-Integral Representation}
It is well-known that integrals over phase-space are, in some situations, not suitable for numerical evaluations
of Wigner functions \cite{moya}.
In this section we obtain a non-integral representation of $W$, in terms of derivatives of Bargmann functions, from expansions around $w=0$. The same argumentation used in the previous section is valid here:
$f^*(z+w/2)f(z-w/2)$ falls-off in a Gaussian-like way for $w \ne 0$.
Let us write
\begin{equation}
[f(z+w/2)]^*=\sum_{n=0}^{\infty}\frac{1}{n!} \frac{{\rm d}^n f^*}{{\rm d} {z^*}^n}\; \left( \frac{w^*}{2}\right)^n\; , \; \; f(z-w/2)=\sum_{j=0}^{\infty}\frac{1}{j!} \frac{{\rm d}^j f}{{\rm d} {z}^j}\; \left( -\frac{w}{2}\right)^j\; .
\end {equation}
Substituting in (\ref{wigsim}) we have
\begin{equation}
\label{wigsim2}
W(z,z^*)=\frac{e^{-z^*z}}{4 \pi \hslash} \sum_{n,j}\frac{(-1)^j}{n!j!} \; \frac{{\rm d}^n f^*}{{\rm d} {z^*}^n} \; \frac{{\rm d}^j f}{{\rm d} {z}^j} 2^{-n-j} \; I_{n,j} \;,
\end{equation}
where
\begin{equation}
I_{n,j}=\int \frac{{\rm d}^2 w}{\pi} \,{w^*}^n w^j\, e^{-w^*w/4+z^*w/2-zw^*/2} \; .
\end{equation}
The integration gives
\begin{eqnarray}
\label{int}
\nonumber
I_{n,j}=4 (-1)^n 2^{n+j} \frac{\partial^n }{\partial z^n}\frac{\partial^j }{\partial {z^*}^j} e^{-z^*z}=
4  (-2)^{n+j} \frac{\partial^n }{\partial z^n}\left(z^j\, e^{-z^*z} \right)\\ \nonumber
=4 (-1)^j 2^{n+j} n! j! {z^*}^n z^j e^{-z^*z} \sum_{\sigma=0}^{n}\frac{(-1)^\sigma}{\sigma ! (n-\sigma)! (j- \sigma)!} |z|^{-2 \sigma}\\ \nonumber
=4 (-1)^j 2^{n+j} n! j! {z^*}^n z^j e^{-z^*z}\; {_2F_0}(-n,-j,-|z|^{-2})\; ,
\end{eqnarray}
where, again, the summation in $\sigma$ may be extended to $\infty$ by means of the already referred property of the Gamma function and, $_2F_0$ denotes the confluent Hypergeometric function of second kind \cite{abramow,grad}. Going back to (\ref{wigsim2}) we obtain the compact expression
\begin{equation}
\label{wighy}
W(z,z^*)=\frac{e^{-2z^*z}}{\pi \hslash} \sum_{n,j} \; _2F_0(-n,-j,-|z|^{-2}) {z^*}^n z^j \frac{{\rm d}^n f^*}{{\rm d} {z^*}^n} \; \frac{{\rm d}^j f}{{\rm d} {z}^j}\; .
\end{equation}
As a simple application we take a harmonic oscillator eigenstate $| \psi \rangle= | N \rangle$. We then have
$f(z)=z^N/\sqrt{N!}$, leading to
\begin{equation}
\nonumber 
W(z,z^*)=N!\, |z|^{2N} \,\frac{e^{-2z^*z}}{\pi \hslash} \sum_{n,j}\frac{_2F_0(-n,-j,-|z|^{-2})}{(N-j)!(N-n)!}\; .
\end{equation}
Summation in $n$ and $j$ readily gives the expected result
\begin{equation}
\nonumber 
W(z,z^*)=N! \,|2z|^{2N} \,\frac{e^{-2z^*z}}{\pi \hslash} \sum_{\sigma}\frac{(-1)^{\sigma}(4z^*z)^{-\sigma}}{\sigma !(N-\sigma)!(N-\sigma)!} = (-1)^N\, \frac{e^{-2z^*z}}{\pi \hslash} L_N(4z^*z)\; ,
\end{equation}
where $L_N$ denotes the Laguerre polynomial of degree $N$.

We note that equation (\ref{wighy}) can be written more elegantly as the quadratic form
\begin{equation}
W(z,z^*)= \frac{e^{-2z^*z}}{\pi \hslash}\,{\bf {V}}^\dagger{\bf {F}}{\bf {V}}\; ,
\label{wigmatrix}
\end{equation}
with the vector ${\bf V}$ and the Hermitian matrix ${\bf F}$ given by
\begin{equation}
{\bf V} \equiv \left(
\begin{array}{c} 
f \\ 
\frac{{\rm d} f}{{\rm d} {z}} \\ 
\frac{{\rm d}^2 f}{{\rm d} {z}^2}\\ 
\vdots
\end{array} \right), \; \;
{\bf F} \equiv \left(
\begin{array}{cccc} 
{_2F_0}(0,0) & {_2F_0}(0,-1)\,z & {_2F_0}(0,-2)\, z^2 & \hdots \\
{_2F_0}(-1,0)\, z^* & {_2F_0}(-1,-1)\, z^*z & {_2F_0}(-1,-2)\, z^*z^2 & \hdots \\
{_2F_0}(-2,0)\, {z^*}^2 & {_2F_0}(-2,-1)\, {z^*}^2z & {_2F_0}(-2,-2)\, {z^*}^2z^2 & \hdots \\
\vdots & \vdots & \vdots & \ddots \\
\end{array} \right) \; ,
\end{equation}
where we have suppressed the third argument $(-|z|^{-2})$ in the matrix elements. Although expression (\ref{wigmatrix}) does not seem to be particularly usefull in simple analytical calculations, it has potential value from both, formal and computational points of view. Equation (\ref{wigmatrix}) may be suitable for numerical evaluations of $W$ in situations where the Bargmann function and its derivatives are easier to obtain than the equivalent integrations in phase space or configuration space (which seems to be usually true). The matrix ${\bf F}$ is built with tabulated functions and the vector ${\bf V}$ has a simple recursive nature, namely ${\bf V}_{j+1}={\rm d} {\bf V}_{j}/{\rm d} {z}$. The diagonal elements of ${\bf F}$ can be put in a simpler form: ${\bf F}_{n,n}=\frac{(-1)^n}{n!}L_n(z^*z)$, which makes clear that the previous definitions of ${\bf F}$ and ${\bf V}$ are less effective in regions of large $|z|$. The alternative definitions ${\bf \tilde{V}}^\dagger=(f,\, z\, {\rm d} f/{\rm d} {z},\, z^2\, {\rm d}^2 f/{\rm d} {z}^2,...)$ and ${\bf \tilde{F}}_{n,j}={_2F_0}(-n,-j,-|z|^{-2})$ (orthogonal matrix) with $W=\frac{e^{-2z^*z}}{\pi \hslash}{\bf \tilde{V}}^\dagger {\bf \tilde{F}}{\bf \tilde{V}}$ may be more convenient, depending on the system under study, for regions of large $|z|$, where the entries of ${\bf \tilde{F}}$ and the derivatives of the Bargmann function are both small.  
\section{$W$ does not depend on $b$: Geometrical Meaning}

A final note on the use of variables $z$ and $z^*$ to express $W$ is in order. Since in the original definition
(\ref{wig0}) there is no mention to the parameter $b$, $W$ must not depend on it regardless the variables we are using to cover the phase-space. Thus, we have d$W/$d$b=0$, which leads to
\begin{equation}
\label{geom1}
b\frac{\partial W}{\partial b}=z^*\frac{\partial W}{\partial z}+z\frac{\partial W}{\partial z^*} \; , 
\end{equation}
where we have used d$z/$d$b=-z^*/b$ and d$z^*/$d$b=-z/b$. It is clear that some explicit dependence on $b$ is required
in order to compensate the implicit dependence contained in $z$ and $z^*$, so that $\partial W/\partial b \ne 0$. As we shall see, the above relation contains a piece of geometrical information. By writing
\begin{equation}
\frac{\partial }{\partial z} = \frac{1}{\sqrt{2}}\left( b\frac{\partial }{\partial q}-i\frac{\hslash}{b}\frac{\partial }{\partial p}\right)\;, \;\; \frac{\partial }{\partial z^*} = \frac{1}{\sqrt{2}}\left( b\frac{\partial }{\partial q}+i\frac{\hslash}{b}\frac{\partial }{\partial p}\right)\;,
\end{equation}
it is easy to show that both sides of equation (\ref{geom1}) are independent of b. We obtain 
\begin{equation}
\label{geom2}
b\frac{\partial W}{\partial b}=q\frac{\partial W}{\partial q}-p\frac{\partial W}{\partial p} \; . 
\end{equation}
This relation can be put in a more suggestive form:
\begin{equation}
\label{geom3}
\frac{b}{\sqrt{q^2+p^2}}\,\frac{\partial W}{\partial b}= {\bf n}\, . \, {\bf grad}\, W \; , 
\end{equation}
where ${\bf n}$ is the unit vector in phase-space parallel to $(q,-p)$ and ${\bf grad}=(\partial /\partial q,\, \partial /\partial p)$ is the gradient operator in phase-space. We see, therefore, that $b \, \partial W/\partial b$ is proportional to the component of the gradient of the Wigner function in the direction $(q,-p)$, i. e., to the variation of $W$ in the ${\bf n}$ direction (see figure 1). 

As an illustration let us assume that $|\psi \rangle$ is itself a coherent state $|U \rangle$, with $\sqrt{2}\,U=(Q/B+iBP/\hslash)$. By using (\ref{wigsim}) (note that $b$ and $B$ not necessarily coincide) it
is easy to obtain
\begin{eqnarray}
\nonumber
W(z,z^*)=\frac{1}{\pi \hslash} \exp \left\{ \frac{B^4-b^4}{2B^2b^2}(z^2+{z^*}^2)- \frac{B^4+b^4}{B^2b^2}z^*z \right.\\ \nonumber
\left.+\frac{B^2-b^2}{Bb}(zU+z^*U^*)+\frac{B^2+b^2}{Bb}(zU^*+z^*U)-2U^*U  \right\} \; .
\end{eqnarray}
It is clear from the corresponding expression in the $(q,p)$ variables, $W(q,p)=\frac{1}{\pi \hslash }\exp \{ -(q-Q)^2/B^2-B^2(p-P)^2/\hslash^2\}$, that $W$ is independent of $b$.
In this particular case, relation (\ref{geom3}) estates that
$b\,\partial W/\partial b = 2W \left[ -\frac{q}{B^2}(q-Q)+\frac{B^2P}{\hslash^2}(p-P)\right]=(q,-p)\, . \, {\bf grad}\, W$ ,  
for any $b$.
\section{Concluding Remarks}
In this work we presented a simple and intuitive derivation of the Bargmann integral representation of Wigner function along with a non-integral representation. We hope that expression (\ref{wigmatrix}) will be of use in numerical computations for a class of theoretical and experimental problems in which the derivatives of the Bargmann functions are readily accessible or, at least, easier to calculate than the integrations in configuration space, equation (\ref{wig0}), or phase-space, equation (\ref{wigsim}). Finally we call attention to the fact that, despite the fact that $W$ does not have an overall dependence on the parameter $b$, its explicit dependence has a simple geometrical meaning.


\acknowledgments
This work was partially supported by the Brazilian agencies CNPq and FACEPE (DCR 0029-1.05/06).

\end{document}